\documentclass[pre,twocolumn,showpacs]{revtex4}
\usepackage{graphicx,latexsym}

\newcommand{\species}[1]{{\itshape #1}}
\newcommand{\abbrspecies}[2]{{\itshape #1.\ #2}}

\newcommand{\Ecoli}{\species{Escherichia coli}}
\newcommand{\Mbark}{\species{Methanosarcina barkeri}}
\newcommand{\Yeast}{\species{Saccharomyces cerevisiae}}

\newcommand{\ecoli}{\abbrspecies{E}{coli}}

\newcommand{\insilicomodel}{i}
\newcommand{\iJR}{\insilicomodel JR}
\newcommand{\iAF}{\insilicomodel AF}
\newcommand{\iND}{\insilicomodel ND}

\newcommand{\latin}[1]{\textit{#1}}

\newcommand{\ie}{\latin{i.\,e.}}

\newcommand{\ya}{y_\alpha}
\newcommand{\va}{v_\alpha}

\newcommand{\vamax}{\va^{\mathrm{max}}}
\newcommand{\vamin}{\va^{\mathrm{min}}}
\newcommand{\vimax}{v_i^{\mathrm{max}}}
\newcommand{\vafix}{\va^{\mathrm{fix}}}

\newcommand{\ve}{v_\energy}
\newcommand{\vemin}{\ve^{\mathrm{min}}}
\newcommand{\vbio}{\mu}
\newcommand{\Sia}{S_{i\alpha}}
\newcommand{\Jia}{J_{i\alpha}}
\newcommand{\Qia}{Q_{i\alpha}}

\newcommand{\Ba}{B_\alpha}

\newcommand{\conc}{c}
\newcommand{\conci}{\conc_i}
\newcommand{\coeff}{b}
\newcommand{\ci}{\coeff_i}
\newcommand{\cenergy}{\coeff_{\energy}}

\newcommand{\qi}{q_i}
\newcommand{\shpr}{\pi}
\newcommand{\mi}{\shpr_i}

\newcommand{\mbio}{\shpr_\biomass}
\newcommand{\exchange}{\rightleftharpoons\emptyset}

\newcommand{\biomass}{{\cal B}}
\newcommand{\energy}{{\cal E}}
\newcommand{\Mi}{\mathrm{M}_i}

\newcommand{\fref}[1]{Fig.~\ref{#1}}
\newcommand{\eref}[1]{Eq.~\ref{#1}}
\newcommand{\sref}[1]{section \ref{#1}}
\newcommand{\opendiamond}{$\Diamond$}
\newcommand{\opensquare}{$\Box$}
\newcommand{\ack}{{\begin{center}---oOo---\end{center}}}
\renewcommand{\subsubsection}[1]{{\itshape #1} : }

\begin{document}

\title{Flux networks in metabolic 
graphs\footnote{Published in Phys. Biol. {\bf6}, 046006 (2009).}}

\author{Patrick B. Warren}
\author{Silvo M. Duarte Queiros}
\author{Janette L. Jones}
\affiliation{Unilever R\&D Port Sunlight, Bebington, 
Wirral, CH63 3JW, UK.}


\begin{abstract}
A metabolic model can be represented as bipartite graph comprising
linked reaction and metabolite nodes.  Here it is shown how a network
of conserved fluxes can be assigned to the edges of such a graph by
combining the reaction fluxes with a conserved metabolite property
such as molecular weight.  A similar flux network can be constructed
by combining the primal and dual solutions to the linear programming
problem that typically arises in constraint-based modelling.  Such
constructions may help with the visualisation of flux distributions in
complex metabolic networks.  The analysis also explains the strong
correlation observed between metabolite shadow prices (the dual linear
programming variables) and conserved metabolite properties.  The
methods were applied to recent metabolic models for \Ecoli, \Yeast,
and \Mbark.  Detailed results are reported for \ecoli; similar results
were found for the other organisms.
\end{abstract}

\pacs{87.16.Yc, 87.18.Vf}


\maketitle

%
\fbox{\centering\begin{minipage}{3.0in}\small
ABBREVIATIONS\\[3pt]
CBM: constraint-based modelling.\\[3pt]
CS: complementary slackness (a property of LP solution pairs at 
optimality).\\[3pt]
GAM: growth-associated maintenance (in relation to ATP consumption).\\[3pt]
gDW: gram dry weight (referring to biomass).\\[3pt]
LP: linear programming.\\[3pt]
NGAM: non-growth-associated maintenance (in relation to ATP consumption).
\end{minipage}}
\vspace{18pt}

\begin{figure}
\begin{center}
\includegraphics{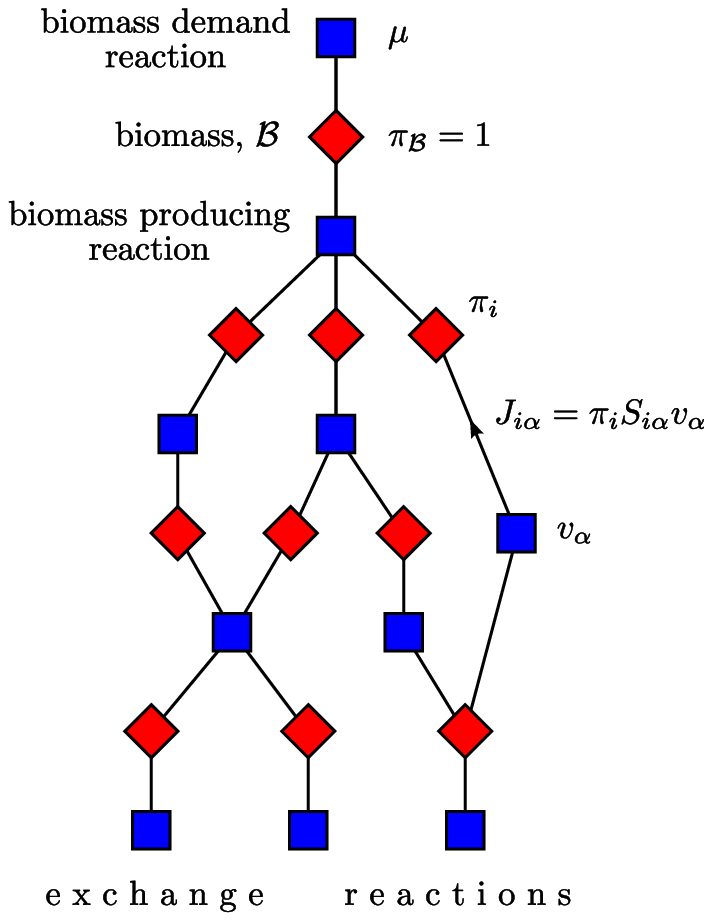}
\end{center}
\caption[]{Schematic metabolic network as a bipartite graph.  Edges go
  from reaction nodes (\opensquare, blue) to metabolite nodes
  (\opendiamond, red).  Reaction fluxes, $\va$, metabolite shadow
  prices, $\mi$, and the stoichiometry matrix, $\Sia$, combine to make
  the edge-associated yield flux network, $\Jia$.  It can be shown
  (see text) that the $\Jia$ are conserved at all nodes except a
  limited number of exchange reactions which serve as sources, and the
  biomass demand reaction which serves as a sink.  The biomass shadow
  price is unity, and the flux through the biomass demand reaction is
  the specific growth rate $\mu$.\label{fig:schem}}
\end{figure}

%
A metabolic network comprises a list of biochemical reactions and
their associated metabolites \cite{cbmbook}.  As such, a convenient
representation is in terms of a bipartite graph containing reaction
nodes and metabolite nodes, with edges between nodes indicating that a
given metabolite is involved in a given reaction \cite{crnbook}.  A
schematic example is shown in \fref{fig:schem}.  The metabolic network
can be modelled by chemical rate equations, giving the rate of change
of the metabolite concentrations in terms of the fluxes, or
velocities, of the associated reactions.  It is widely accepted though
that the metabolism comes to a steady state very quickly, so that the
metabolite concentrations are unchanging in time.  This means that a
flux balance condition holds, and the set of reaction fluxes (the
`fluxome') can, essentially, be regarded as the metabolic phenotype.
Determination of the fluxome is therefore the focus of
considerable theoretical \cite{cbmbook}, and experimental effort
\cite{Sauer04, SKS}.  The global properties of such flux sets have
been investigated \cite{barabasi}.

When the network is represented as a bipartite graph, the fluxome is
traditionally associated with the reaction nodes.  Here we show how
fluxes can be associated with the \emph{edges} of such a bipartite
graph, by combining the reaction fluxes with any metabolite property
that is conserved in the majority of reactions, such as molecular
weight.  Moreover, assuming the flux balance condition, such an
edge-associated flux network is conserved at all the reaction and
metabolite nodes apart from a handful of sources and sinks.  Thus the
edge-associated fluxes resemble, for example, electric currents in a
network of resistors \cite{GSHM07}.  This observation may help with
the visualisation of the flow of material in these complex reaction
networks.  If the flux-balance condition does not hold (for example
away from steady state), then the edge-associated flux network can
still be constructed provided a set of reaction fluxes is available.
In such a case though, the edge fluxes are not in general conserved at
the reaction nodes. Finally in the case where the set of reaction
fluxes arises from the solution to a linear optimisation or linear
programming (LP) problem, such as commonly encountered in
constraint-based modelling (CBM), then an edge-associated \emph{yield
  flux network} can be constructed.

CBM is now a well-established approach for calculating
candidate sets of reaction fluxes \cite{cbmbook}.  It has been applied
to micro-organisms from all three domains of life \cite{iJR904,
  iND750, iAF692, iAF1260}, and recently extended to encompass human
metabolism \cite{human}.  For the growth of micro-organisms a commonly
used paradigm has emerged in which the metabolic network is augmented
with a biomass reaction consuming the end-points of metabolism in the
appropriate ratios, and with exchange reactions to represent the
uptake of substrates and the discharge of metabolic by-products.
Maximising the flux through the biomass reaction amounts to maximising
the specific growth rate of the micro-organism.  This approach has
been highly successful at predicting the behaviour of micro-organisms
\cite{IEP, TS, SKS}, and has also been applied to problems in
metabolic engineering \cite{Zhang, BPM03}.  

As already indicated CBM typically leads to an LP problem for the set
of reaction fluxes.  Mathematically, every LP problem has an
associated dual \cite{lpbook, numrec3}.  The dual variables are known
as shadow prices reflecting an economic interpretation of the dual
problem.  In CBM, shadow prices were first investigated by Varma and
Palsson who showed that they are, essentially, yield coefficients
\cite{VP1, VP2}.  Later, the dual problem was explicitly formulated by
Burgard, Maranas, and coworkers, for use in multi-level optimisation
problems \cite{BM03, BPM03}.  Recently we described a thermodynamic
interpretation of the dual problem \cite{WJ07}.  The aforementioned
yield flux network can be generated very naturally by combining a
candidate set of reaction fluxes with the corresponding set of shadow
prices.  The properties of the yield flux network (such as flux
conservation) follow from the so-called complementary slackness (CS)
relations.  The parallel construction of the various flux networks
explains the strong correlation between shadow prices and
conserved metabolite quantities such as molecular weight.

\section{Methods}
\subsection{Conserved edge-associated flux networks}\label{sec:eafn}
Mathematically, a metabolic network is conveniently described by the
stoichiometry matrix $\Sia$, giving the number of moles of the $i$-th
metabolite consumed or produced by the $\alpha$-th reaction.  We make
a distinction between balanced `internal' reactions which typically
represent biochemical transformations or membrane transport processes,
and imbalanced reactions introduced in CBM such as the biomass
reaction or the exchange reactions.  We suppose there are $\alpha = 1
\dots R$ reactions (of all types) and $i = 1 \dots M$ metabolites,
with typically $M < R$.  The convention we adopt is that $\Sia$ is
positive for products, negative for reactants.  The corresponding
bipartite graph has $R$ reaction nodes and $M$ metabolite nodes
(\fref{fig:schem}).  An edge connects a reaction node to a metabolite
node if and only if $\Sia\ne0$.  We additionally suppose the bipartite
graph is directed and adopt the convention that all edges start at
reaction nodes and end on metabolite nodes.

In terms of the stoichiometry matrix, the chemical rate equations are
\begin{equation}
\frac{d\conci}{dt} = \textstyle\sum_\alpha \Sia \va
\end{equation}
where $\conci$ is the concentration of the $i$-th metabolite, and $\va$
is the flux through the $\alpha$-th reaction (reaction velocity).  The
reaction fluxes are typically measured in units of $\mathrm{mol} /
\mathrm{gDW.hr}$ where gDW means gram dry weight of biomass.  In
steady state, $d\conci/dt = 0$, leading to the flux balance
condition
\begin{equation}
\textstyle\sum_\alpha \Sia \va = 0.\label{eq:fb}
\end{equation}

Let $\qi$ be any property of the $i$-th metabolite which is conserved
in the internal reactions, for example molecular weight, number of
atoms, and so on.  Then the corresponding edge-associated flux network is
defined by
\begin{equation}
\Qia=\qi\Sia\va\label{eq:qia}
\end{equation}
(no implied summation).  For example if $\qi$ is molecular weight we
generate what we term the mass flux network, if $\qi$ is the number of
atoms of a given element we generate an elemental flux network, and so
on.  From our sign convention for the stoichiometry matrix one has
$\Qia>0$ for edges connected to product metabolite nodes, assuming
$q_i>0$ and $\va>0$.  Hence, typically, the flux flows from reactants
to products.

The flux network $\Qia$ defined in \eref{eq:qia} is conserved at all
the metabolite nodes since
\begin{equation}
\textstyle\sum_\alpha \Qia = 
q_i (\sum_\alpha \Sia \va) = 0,
\end{equation}
using the flux balance condition \eref{eq:fb}.  Moreover $\Qia$ is
conserved at all the internal reaction nodes since
\begin{equation}
\textstyle\sum_i \Qia = 
(\sum_i q_i \Sia\bigr) \va = 0
\end{equation}
where, by definition, $\sum_i q_i \Sia = 0$ expresses the fact that
$q_i$ is conserved in the $\alpha$-th reaction.  However, $\Qia$ may
not necessarily be conserved at imbalanced reaction nodes since
$\sum_i q_i \Sia$ is not necessarily zero.  Clearly these are
important nonetheless, since they represent sources and sinks for the
$\Qia$.  As discussed in more detail below, in a typical metabolic
model these imbalanced reaction nodes are restricted to a small set
of so-called exchange and demand reactions (see \fref{fig:schem}).

In some sense these edge-associated flux networks, and the yield flux
network introduced below, are more fundamental than the set of
reaction fluxes.  By construction they are insensitive to either local
rescaling of the reaction stoichiometry ($\Sia\to\Sia\times r_\alpha$
and $\va\to\va/r_\alpha$), or local coarse-graining of metabolites
($\Sia\to \Sia\times r_i$ and $\qi\to\qi/r_i$).  Since these
rescalings apply locally, like a gauge transformation, one
might say the edge-associated flux networks are gauge invariant.

\subsection{The linear programming problem in constraint-based
  modelling}
As outlined in the introduction, the typical application of CBM leads
to a so-called primal LP problem, with a corresponding dual LP
problem.  The details are discussed in the following subsections.

\subsubsection{Primal linear programming problem}
The variables in the primal LP problem are the reaction fluxes $\va$.
The constraints are the flux balance conditions in \eref{eq:fb},
augmented by additional constraints as follows.  Firstly,
thermodynamic considerations may lead to some of the internal
reactions being judged to be irreversible in which case $\va\ge0$.  In
principle the reaction fluxes can also be `capped'
($\vamin\le\va\le\vamax$) or fixed to a prescribed value
($\va=\vafix$) although in practice this is rarely done.

Next, exchange / demand reactions represent the uptake or discharge of
substrates from the environment.  By convention they are imbalanced
reactions of the form $\Mi\exchange$, where $\Mi$ is a metabolite, so
a positive flux represents discharge of the corresponding substrate
and negative flux represents uptake from the environment.  For these
reactions $\Sia = -1$.  There is no distinction between demand and
exchange reactions, although we tend to restrict the phrase `demand
reaction' to the case when the flux is expected to be positive.
Exchange / demand reactions are classified according to the allowed
flux range, as follows :
\begin{equation}
\va\;\left\{\begin{array}{ll}
= 0 & \mbox{(closed)},\\
\ge 0 & \mbox{(half-closed; uptake prevented)},\\
\ge-\vimax & \mbox{(open to a limited extent)},\\
\mathrm{unconstrained} & \mbox{(fully-open)}.\\
\end{array}\right.
\end{equation}
In the third of these, $\vimax$ is the maximum specific uptake rate of
the given substrate.  Most of exchange reactions are half-closed,
since there is a need to prevent arbitrary uptake.  For certain
essential minerals, dissolved gases, nutrients, and vitamins, the
exchange reactions are fully open so that the corresponding substrates
can be freely taken up or discharged by the organism.  In typical
applications, one or two exchange reactions are also opened to a
limited extent, representing growth-limiting substrates such as
carbon/energy sources.

The biomass reaction consumes the end-points of metabolism such as
amino acids, nucleotides, lipids, and co-factors.  In an extension to
the usual paradigm, we split this into an irreversible biomass
producing reaction $\sum_i \ci \Mi \to \biomass$ and an open biomass
demand reaction $\biomass\exchange$, where $\biomass$ is a artificial
metabolite representing biomass.  The stoichiometry coefficients $\ci$
typically have units $\mathrm{mol}/\mathrm{gDW}$.  The flux through
the biomass demand reaction is the specific growth rate, $\vbio$,
typically with units $1/\mathrm{hr}$.  Since $\biomass$ only features
in these two reactions, the irreversibility of the production step
ensures $\mu\ge0$.

Finally, the energetic requirements of the organism are taken care of
by including growth associated maintenance (GAM) and (for high
accuracy work) non growth associated maintenance (NGAM) reactions.
Again in an extension to the common paradigm, we account for these by
including an irreversible energy-generating reaction
$\mathrm{ATP^{4-}} + \mathrm{H_2O} \to \mathrm{ADP^{3-}} +
\mathrm{H^+} + \mathrm{HPO_4^{2-}} + \energy$ where $\energy$ is a
artificial metabolite representing the energy that can be gained by
hydrolysing (one mole of) ATP.  For the GAM, $\energy$ is included
amongst the metabolites consumed in the biomass producing reaction
with the coefficient $\cenergy$ representing the GAM requirement.  For
the NGAM, we add an energy demand reaction $\energy\exchange$ with a
positive lower bound for the flux, $\ve\ge\vemin$ where $\vemin$
represents the NGAM requirement.

This completes the specification of the primal LP problem in typical
applications of CBM.  The constraints on $\va$ specify a so-called
feasible solution space.  The aim is to maximise the specific growth
rate, $\vbio$, whilst remaining within the feasible solution space.
The introduction of $\biomass$ and $\energy$ allows us to move the
non-trivial flux bounds and the target of the optimisation to demand
reactions, with a corresponding simplification to the dual problem.
Numerically, solutions to this LP problem can be found by a
straightforward application of LP techniques, for example the simplex
algorithm \cite{lpbook, numrec3}.  A MATLAB toolbox for solving LP
problems in CBM has been released by the Palsson group \cite{COBRA}.
For the present study, we used a bespoke interface to the GNU LP kit
(GLPK), which provides an efficient implementation of the simplex
algorithm.

\subsubsection{Dual linear programming problem}\label{subsec:dual}
Now we turn to the dual LP problem.  The dual variables are shadow
prices associated with constraints in the primal LP problem.  In
particular the flux-balance conditions in \eref{eq:fb} generate a set
of shadow prices $\mi$ for the metabolites.  The shadow prices in the
dual problem are then subject to constraints that correspond to the
variables (reaction fluxes) in the primal problem.  As mentioned
above, the shadow prices can be interpreted as yield coefficients
\cite{VP1, VP2}.  A full derivation of the construction rules
presented below is given in the Appendix.

For the internal reactions, the constraints on the $\mi$ can be
written in terms of the derived quantities \cite{WJ07}
\begin{equation}
\textstyle\Ba = \sum_i \mi \Sia.\label{eq:ba}
\end{equation}
These are defined for all reactions (the resemblence to reaction
affinity will be discussed shortly).  In the dual problem, each
reversible or irreversible internal reaction generates a constraint on
the $\Ba$ according to :
\begin{equation}
\Ba\;\left\{\begin{array}{ll}
= 0 & \mbox{(reversible, $\va$ unconstrained)},\\
\le 0 & \mbox{(irreversible, $\va\ge0$)}.
\end{array}\right.\label{eq:ba2}
\end{equation}
This result applies only in the two cases indicated.  The
generalisation to more complicated situations, such as fixed fluxes
and double-bounded fluxes, is given in the Appendix.

The constraints arising from the internal reactions are supplemented
by additional constraints arising from the exchange and demand
reactions.  Since these reactions involve only one metabolite
($\Mi\exchange$ with $\Sia = -1$), the corresponding constraint
simplifies to feature only the corresponding metabolite shadow price.
The constraints associated with the exchange / demand reactions are :
\begin{equation}
\mi\;\left\{\begin{array}{ll}
= 0 & \mbox{(fully open, $\va$ unconstrained)},\\
\ge 0 & \mbox{(half-closed, $\va\ge0$},\\
 & \mbox{{}\qquad or limited, $\va\ge-\vimax$)},\\
\mathrm{unconstrained} & \mbox{(closed, $\va=0$)}.
\end{array}\right.
\end{equation}
The biomass demand reaction is special and the corresponding
constraint is 
\begin{equation}
\mbio = 1\quad\mbox{(biomass)}.
\end{equation}
This arises because it is the flux through the biomass demand reaction
that is the optimisation target in the primal LP problem.  It also
make sense since by definition the biomass yield coefficient for
adding more biomass is unity.  Given that $\mbio = 1$ is
dimensionless, the units of $\mi$ for the other metabolites are
typically $\mathrm{gDW}/\mathrm{mol}$ (inverse to the units of $\ci$).
Again this makes sense in terms of the $\mi$ being yield coefficients.

The objective function in the dual LP problem is to minimise $w =
\sum_i \mi\vimax$, where the sum is over the limited exchange
reactions only.  One can show from the strong duality theorem
\cite{lpbook} that the minimum value of $w$ is equal to the maximum
value of the specific growth rate $\vbio$, provided both problems have
solutions.  It is quite common that there is only one limited exchange
reaction, representing single-substrate limitation.  In this case the
dual objective can be taken to minimise the shadow price of the
corresponding metabolite.  It follows that $\vimax$ does not enter the
dual problem any more, and the metabolite shadow prices are
independent of growth rate.  Also, at optimality one has $\vbio = w =
\mi\vimax$.  But $\vbio/\vimax$ is the standard definition of the
yield coefficient, confirming the interpretation of $\mi$ as the yield
coefficient for the limiting substrate.

This construction of the dual LP problem extends and simplifies the
results presented in \cite{WJ07}.  It also essentially
recovers the results obtained by Burgard and Maranas and coworkers
\cite{BM03, BPM03}.  Numerically, the dual problem can of
course be solved directly, however the shadow prices are often
obtained `for free' as a by-product of the primal LP solution method.
This is the case with the simplex algorithm for instance
\cite{lpbook}.

\subsubsection{Complementary slackness relations}
At optimality a number of so-called complementary slackness (CS)
relations hold, linking the solutions to the primal and dual LP
problems.  These are also derived in the Appendix.  For irreversible
internal reactions the CS relation is $\Ba\va=0$.  There is no CS
relation for reversible internal reactions but, since $\Ba=0$ is
imposed, it follows that at optimality $\Ba\va=0$ for all internal
reactions.  For the exchange reactions the CS relations are $\mi\va=0$
for half-closed exchange reactions and $\mi(\va+\vimax)=0$ for the
limited exchange reactions.  There is no CS relation for fully open
exchange reactions but again, since $\mi = 0$ is imposed (apart from
biomass), it follows that at optimality $\mi\va = 0$ holds for all
exchange reactions except for limited exchange reactions operating at
the lower flux bound ($\va=-\vimax$).  There is no CS relation for the
biomass demand reaction since it is assumed fully open.

\subsection{The yield flux network}\label{sec:yfn}
We now show how a pair of complementary solutions to the above primal
and dual LP problems can be used to construct a naturally conserved
edge-associated flux network, similar to those constructed in
\sref{sec:eafn}.  We start by defining the quantities
\begin{equation}
\Jia = \mi\Sia\va\label{eq:jia}
\end{equation}
which typically have units of $1/\mathrm{hr}$. The $\Jia$ are
conserved at metabolite nodes since  
\begin{equation}
\textstyle
\sum_\alpha\Jia = \mi(\sum_\alpha\Sia\va) = 0
\end{equation}
follows from the flux balance condition.  
At optimality the $\Jia$ are also conserved at internal
reaction nodes since
\begin{equation}
\textstyle \sum_i\Jia = (\sum_i\mi\Sia)\va = \Ba\va = 0,
\end{equation}
from complementary slackness.  

Exchange reaction nodes are linked by a single edge to the
corresponding metabolite.  For these edges, $\Sia = -1$ and $\Jia =
-\mi\va$.  Complementary slackness therefore implies $\Jia = 0$ unless
the exchange reaction happens to be operating at the lower flux bound
in which case $\Jia = \mi\vimax$.  For the biomass demand reaction
one has $\Jia = -\vbio$ since $\va = \vbio$ and $\mbio = 1$.

Thus we conclude that at optimality the yield fluxes $\Jia$ are
conserved at all nodes of the bipartite graph, except for exchange
reaction nodes where the reaction happens to be operating at the lower
flux bound, which act as sources, and the biomass demand reaction
node, which acts as a sink.  We argue this justifies the notion that
the $\Jia$ constitute a `yield flux network' indicating how material
which contributes to growth is transmitted through the network.  If
there is only one limited exchange reaction, the conservation law
derived above implies $\mi\vimax = \vbio$.  But at optimality the
strong duality theorem shows this is true, as we have discussed in
\sref{subsec:dual}.

\subsection{Comparison between flux networks}
The yield flux network $\Jia$ and the flux networks $\Qia$ introduced
in \sref{sec:eafn} are very similar.  They share the \emph{same} small
subset of reaction nodes which act as sources and sinks, namely the
exchange and demand reaction nodes (\fref{fig:schem}).  Thus one might
expect the pattern of fluxes to be similar.  Moreover, since $\Jia$
and $\Qia$ are constructed from identical stoichiometry coefficients
and reaction flux sets (see \eref{eq:qia} and \eref{eq:jia}), this
suggests that one would expect a strong correlation between the shadow
prices $\mi$ and conserved metabolite properties $\qi$.  This is
confirmed by studies of genome-scale metabolic reconstructions,
discussed in section \ref{sec:res}.  We cannot provide a proof of an
exact relationship between the $\mi$ and $\qi$, and indeed the
correlation is not expected to be perfectly linear since the
corresponding source terms are not necessarily in exact proportion.
For example the yield flux network will typically have a single
exchange reaction node as a source node (\ie\ the one operating at the
lower flux bound), whereas the mass flux network has sources at all
the exchange reaction nodes which carry a reaction flux (since all
metabolites have some molecular weight).  By the same argument, it is
of course not surprising that different conserved molecular quantities
are only approximately linearly correlated, for example, molecular
weight is only approximately proportional to atom count.

\subsection{Analogies to chemical thermodynamics}\label{sec:thermo}
Recently we described a thermodynamic interpretation of the dual
problem \cite{WJ07}.  For completeness, we summarise the analogies and
differences between the dual LP problem and non-equilibrium
thermodynamics as applied to these networks by Beard and coworkers
\cite{Beard1, Beard2, Beard3}.  There is obviously an analogy between
\eref{eq:ba} and \eref{eq:ba2}, and conventional chemical
thermodynamics \cite{smithbook}, wherein
\begin{equation}
\begin{array}{lcl}
\mi & \leftrightarrow & \mbox{chemical potential},\\
\Ba & \leftrightarrow & \mbox{reaction affinity}.
\end{array}
\end{equation}
However the analogy fails to be exact since the CS conditions require,
at optimality, $\Ba\va=0$.  This means that whenever there is a flux
through a reaction ($\va\ne0$) the corresponding `affinity' vanishes
($\Ba=0$).  This stands in sharp contrast to conventional
non-equilibrium thermodynamics where a flux through a reaction is
usually associated with a negative reaction affinity (driving force)
\cite{Beard1, Beard2, Beard3}.  It shows that the thermodynamic
interpretation of the dual problem cannot be put into exact
correspondence with conventional non-equilibrium thermodynamics.

\subsection{Genome-scale metabolic reconstructions}\label{sec:gsmr}
A genome-scale metabolic reconstruction encompasses all the
biochemical transformations allowed for by enzymes encoded on the
genome of the organism of interest.  As such it represents the entire
metabolic capability of the organism.  A growing number of such
reconstructions are becoming available.  The principal genome-scale
model used in the present study is \iAF1260\ for
\Ecoli\ \cite{iAF1260}, which is perhaps the most complete
whole-organism model currently available.  We have also studied
\iND750 for \Yeast\ \cite{iND750}, and \iAF692 for the archeal
methanogen \Mbark\ \cite{iAF692}, in addition to an earlier model
\iJR904 for \ecoli\ adjusted slightly to account for later literature
\cite{iJR904, RP2, KPH}.  

The energetic requirements in these models are represented by a GAM
component in the biomass producing reaction and a separate NGAM
reaction.  For our calculations we retain the GAM requirement, but the
NGAM requirement was turned off for simplicity ($\vemin=0$).  We have
checked that this approximation has little influence on our results.
Exchange reactions are provided for all the extracellular metabolites in
these models.  By default they are half-closed, meaning the flux is
constrained to be non-negative so discharge only is possible.  A
subset of the exchange reactions are made fully open, on a
case-by-case basis (details available on request).  In our
calculations \emph{one} additional exchange reaction was also opened
to a limited extent representing the limited availability of a
substrate under single-substrate growth limitation conditions.

For genome-scale problems it is often the case that even at optimality
the flux distribution is still not uniquely constrained, because there
may be alternate pathways in the metabolism.  Mathematically this
shows up in the existence of alternate optima in the LP problem
\cite{MS}.  This is an interesting phenomenon which somewhat
complicates the analysis.  If the primal LP problem has alternate
optima, there will be a similar plurality of solutions to the dual
problem.  However for every optimum solution to the primal problem,
there exists a complementary optimal solution to the dual problem.
For instance the shadow prices generated `for free' by the simplex
algorithm are automatically complementary to the primal solution.  It
is important to note that the yield flux network described above must
be constructed using complementary solution pairs, since the CS
relations apply only in this case.  The genome-scale models discussed
below exhibit the phenomenon of alternate optima.  However we have
checked rather carefully that we have used representative solution
pairs in presenting our results.

\subsection{Statistical analysis}\label{sec:stats}
\begin{figure}
\begin{center}
\includegraphics{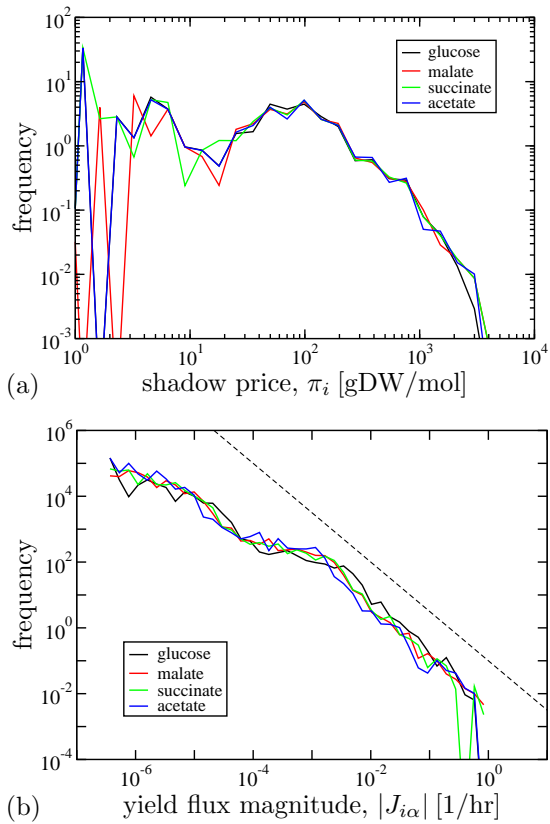}
\end{center}
\caption[]{Distribution of (a) shadow prices and (b) yield fluxes for
  \ecoli\ (\iAF1260) growing on various substrates under aerobic
  conditions.  The dashed line in (b) is the power-law
  $|\Jia|^{-3/2}$.\label{fig:distro}}
\end{figure}

We undertook statistical analysis of the shadow price distributions
for selected conditions and organisms although the results are
somewhat inconclusive.  We attempted to fit the observed
distributions, using maximum likelihood estimators, to a log-Normal, a
$\chi$ distribution, an inverted $\chi$ distribution, and a general
distribution $\sim\exp[-(\mi^2+\omega\beta)/\omega \mi]$ which has
tunable exponential asymptotic behaviour for large and small $\mi$
($\omega$ and $\beta$ are parameters).  However none of these
distributions could be said to fit the observed distributions, as
judged by the Kolmogorov-Smirnov test \cite{biostatbook}.  Aside from
attempting to fit the full distribution, we also determined whether
the observed distributions were compatible with power-law asymptotic
behaviour at large and small $\mi$, using tail estimators of Hill
\cite{Hillstat} and Meerschaert-Scheffer \cite{MSstat}.  Our results
gave exponents systematically greater than 3 (see for example
\fref{fig:distro}, upper plot), which is generally taken to be an
indication of exponential asymptotic behaviour rather than power-law
behaviour.  When we apply these tail estimators to the distribution
for $|\Jia|$, we recover an exponent value $\approx -3/2$ for large
magnitudes, shown as the dashed line in figure \ref{fig:distro}(b).

\begin{figure}
\begin{center}
\includegraphics{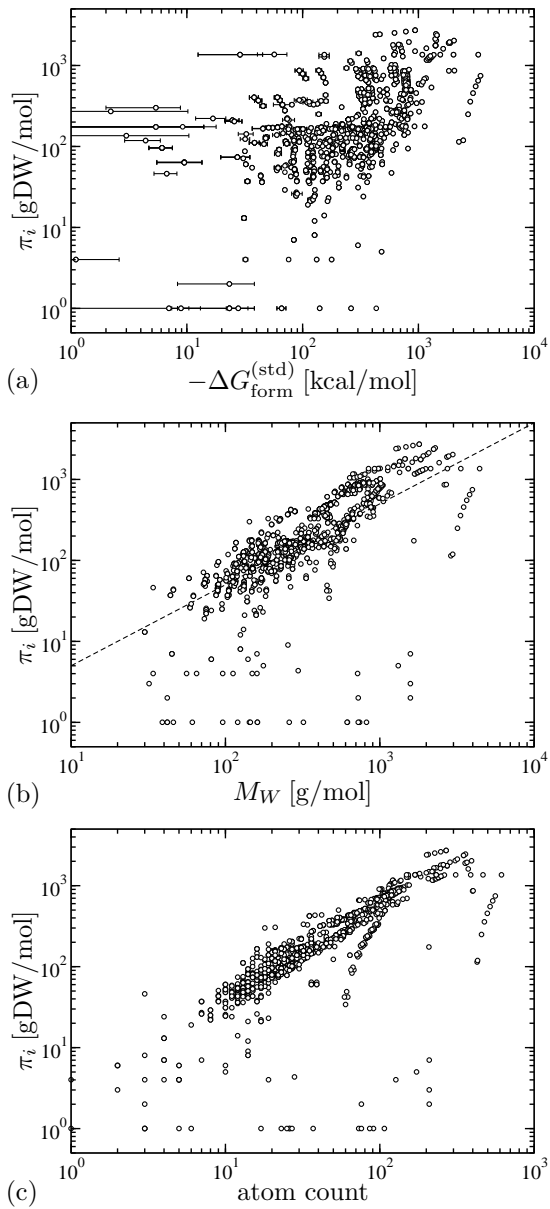}
\end{center}
\caption[]{Shadow prices for \ecoli\ (\iAF1260) growing on glucose
  under aerobic conditions as a function of (a) (minus) metabolite
  formation free energy, (b) molecular weight, and (c) total atom
  count.  The dashed line in (b) is $\mi=0.5\times M_W$ corresponding
  to a mass yield coefficient of
  $0.5\,\mathrm{gDW}/\mathrm{g}$.\label{fig:comp}}
\end{figure}

\begin{figure}
\begin{center}
\includegraphics{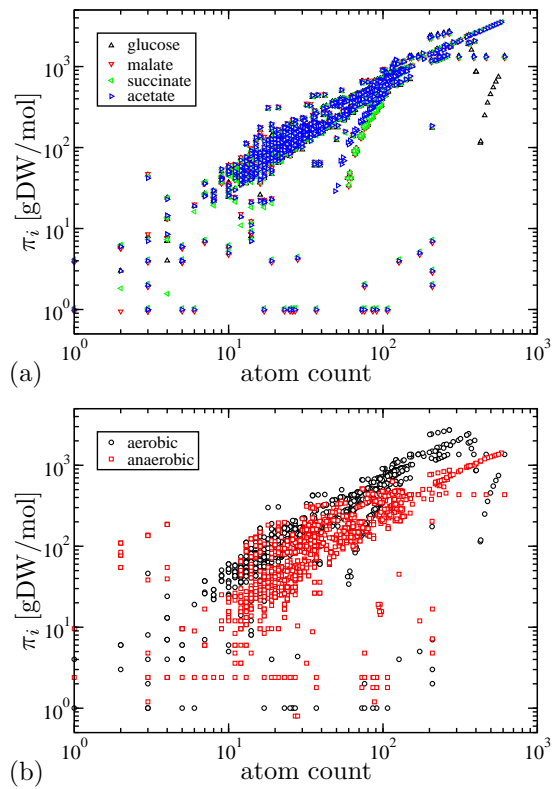}
\end{center}
\caption[]{Shadow prices for \ecoli\ (\iAF1260) growing (a) on various
  substrates under aerobic conditions, and (b) on glucose under
  aerobic and anaerobic conditions.\label{fig:growth}}
\end{figure}

\section{Results and discussion}\label{sec:res}
We report results for \iAF1260 for \ecoli\ \cite{iAF1260}, under
various conditions.  Similar results were obtained for the other
organisms and models studied.  The data used to generate figures
\ref{fig:distro}--\ref{fig:growth} has been compiled into an Excel
spreadsheet, and is given as supplementary material.

We first discuss the statistical distribution of the shadow prices and
yield fluxes.  Figure \ref{fig:distro}(a) shows that the shadow prices
in \iAF1260\ have a broad distribution of around three orders of
magnitude.  We used statistical tests described in \sref{sec:stats} to
analyse the distribution, however these were rather inconclusive.  We
have concluded though that there is unlikely to be any asymptotic
power-law behaviour in the distributions.  For the cases studied,
$\approx 80$\% or more of the metabolites have a positive shadow
price, $\approx 15$\% have a zero shadow price meaning that the growth
rate is unchanged if the provision of these metabolites is altered,
and $\approx 5$\% or less have a negative shadow price meaning the
growth rate actually goes down if that metabolite is injected into the
system.  Figure \ref{fig:distro}(b) shows the distribution of the
yield fluxes $\Jia = \mi\Sia\va$.  This distribution \emph{does}
appear to show asymptotic power-law behaviour for large magnitudes.
It is notable that the exponent appears to be the same as has been
found for the reaction flux distribution \cite{barabasi}.  This is
interesting since the yield flux network is gauge invariant in the
sense discussed at the end of section \ref{sec:eafn}, whereas the
reaction fluxes are not gauge invariant.  The fact that we observe
power-law behaviour in the yield flux network therefore strengthens
the earlier analysis of \cite{barabasi}.

Now we turn to the correlation between shadow prices and conserved
molecular properties.  Figure \ref{fig:comp} shows the shadow price as
a function of metabolite formation free energy, molecular weight, and
total atom count.  To obtain these plots, metabolite formation free
energies (where available) and molecular weights are taken from
\cite{iAF1260}, and the atom count is computed from the atomic
formulae in \cite{iAF1260}.  The weakest correlation is with (minus)
the free energy of formation.  This is unsurprising since the
formation free energy is imperfectly conserved in reactions.  A
stronger correlation is found with molecular weight and the strongest
correlation is with atom count.  These quantities are conserved since
reactions in these genome-scale models are charge- and mass-balanced.
The shadow price is more strongly correlated with atom count than
molecular weight because there is less spread in the magnitude of the
values for atom count.  The shadow prices discussed here are `molar'
yield coefficients.  One can of course define a `mass' yield
coefficient by dividing by the molecular weight.  The dashed line in
figure \ref{fig:comp}(b) shows that the mass yield coefficients are
approximately constant with a value of $\approx 0.5\, \mathrm{gDW} /
\mathrm{g}$ (we have drawn back from undertaking a linear regression
analysis as we believe this would over-interpret the data and would
not add any new insights).

The use of shadow prices to measure efficiencies in a model of the
central metabolism of \ecoli\ was pioneered by Varma and Palsson
\cite{VP1, VP2}.  Our calculations extend the scope of this analysis
to more recent genome-scale metabolic models.  Figure \ref{fig:distro}
and figure \ref{fig:growth}(a) shows the shadow prices for
\ecoli\ grown on four different limiting carbon/energy sources.
Despite the spread in growth rates from these sources, by and large
there is little difference in the shadow price distribution.  Invoking
the efficiency arguments of Varma and Palsson, this plot suggests that
the metabolic network of \ecoli\ has evolved to be equally efficient
for growing on a variety of substrates.  This reflects the `bow-tie'
structure of the metabolic network \cite{mazeng, TCD, ZYLCL}, as
substrates are first broken down to a dozen or so common precursors,
before being re-assembled into the components required for growth.

Figure \ref{fig:growth}(b) shows a significant overall lowering of the
shadow prices for anaerobic growth compared to aerobic growth.  This
reflects a reduced efficiency of the network, as more effort has to go
into satisfying the energetic requirements of the organism in the
absence of oxidative phosphorylation.  Two further calculations
support this conclusion (data not shown).  Firstly, a similar
reduction in shadow prices is found for growth under aerobic
conditions with the ATP synthase reaction disabled.  Secondly, if the
NGAM energy demand reaction is thrown fully open ($\energy\exchange$)
so that the organism can trivially satisfy its energy requirements,
the shadow prices are practically unchanged on going from aerobic to
anaerobic conditions.

\section{Conclusion and outlook}
To summarise, we have examined the problem of constructing conserved
flux networks defined on the edges of bipartite metabolic graphs.  We
find that such networks can be generated by combining a conserved
metabolic property such as molecular weight, with the reaction fluxes.
A similarly conserved, edge-associated flux network can be constructed
from a natural combination of the primal and dual solutions to the LP
problem that typically arises in CBM.  The correspondence between
these edge-associated flux networks is responsible for the high
correlation between shadow prices and conserved molecular properties.
The construction of these networks opens the way for further
investigations, both of the global properties of the flux distribution
(\fref{fig:distro}) \cite{barabasi}, and the scaling theory of
transport in complex networks \cite{GSHM07}.

\ack

SMDQ acknowledges support by the European Commission through the Marie
Curie Transfer of Knowledge project BRIDGET (MKTD-CD 2005029961)


\appendix
\section*{Appendix}
\setcounter{section}{1}
This appendix presents for completeness a derivation of the dual LP
problem and the CS relations for the typical LP problem that arises in
CBM.  It is based on the development in \S6.5 in \cite{lpbook} (see
also \cite{numrec3}).  In this approach we assign a Lagrange
multiplier to each constraint in the primal problem, for example the
shadow prices are multipliers associated with the flux balance
constraints of \eref{eq:fb}.  Flux constraints are handled by
conversion to quadratic equalities.  Thus, restricting the analysis
for the time being to the two most common flux constraints, we have
\begin{equation}
\begin{array}{lll}
\va\ge0 & \Rightarrow & \va=u_\alpha^2\,,\\
\va\ge-\vimax & \Rightarrow & \va+\vimax = u_\alpha^2\,.
\end{array}
\end{equation}
The first corresponds to irreversible internal reactions and
half-closed exchange reactions.  The second corresponds to limited
exchange reactions.  Adopting the Lagrange multiplier approach, we
replace the original constrained linear optimisation problem by the
following problem in which we seek the unconstrained maximum of
\begin{equation}
\begin{array}{l}
Z=\vbio+\sum_{i\alpha}\mi\Sia\va
+\sum'_\alpha \ya(\va-u_\alpha^2)\\[6pt]
\hspace{6em}{}+\sum''_\alpha \ya(\va+\vimax-u_\alpha^2)
\end{array}
\end{equation}
where the first term is the original objective function $\vbio$, the
second term incorporates the flux balance conditions with $\mi$
being Lagrange multipliers, and the third and fourth terms accommodate
the quadratic equalities with $\ya$ being the multipliers.  The prime
and double prime restrict the sums to the respective reactions with a
zero or non-zero lower flux bound.  We rewrite this as
\begin{equation}
\begin{array}{l}
Z=\sum''_\alpha \ya\vimax+\vbio+\sum_\alpha\Ba\va\\[6pt]
\hspace{4em}{}+\sum'_\alpha \ya\va
+\sum''_\alpha \ya\va\\[6pt]
\hspace{6em}{}-\sum'_\alpha \ya u_\alpha^2
-\sum''_\alpha \ya u_\alpha^2
\end{array}
\label{eq:app1}
\end{equation}
where $\Ba = \sum_i\mi\Sia$ is introduced as a definition to
correspond to the main text.  

From \eref{eq:app1} one condition that $Z$ is an extremum is
${\partial Z}/{\partial\va}=0$, implying
\begin{equation}
\Ba = \left\{
\begin{array}{ll}
0 & (\mathrm{unbound}),\\
-\ya & (\mathrm{bound}),\\
-1 & (\mathrm{biomass})
\end{array}
\right.\label{eq:appa}
\end{equation}
(`unbound' for unbounded reactions; `bound' for reactions with a zero
or non-zero lower flux bound, `biomass' for the biomass demand
reaction which we assume is unbounded).  We notice from
\eref{eq:app1} that $\ya\ge0$ is required for $Z$ to be a
\emph{maximum}, otherwise $Z\to\infty$ as $u_\alpha\to\pm\infty$.
From this and the second of \eref{eq:appa} we deduce that
$\Ba\le0$ for reactions with a lower flux bound.  Taken with the first
of \eref{eq:appa} this gives the $\Ba$-conditions for the
internal reactions quoted in the main text.  For the exchange
reactions, one has $\Sia=-1$ for the metabolite involved in the
reaction, and $\Sia=0$ for every other metabolite.  Hence $\Ba=-\mi$.
From this, and \eref{eq:appa} and $\ya\ge0$, we recover the
conditions on the exchange reaction metabolite shadow prices quoted in
the main text.

The CS relations follow from ${\partial Z}/{\partial u_\alpha} = 0$.
This implies $\ya u_\alpha = 0$, and hence $\Ba\va = 0$, or
$\Ba(\va-\vamin)=0$, for reactions with a zero, or non-zero, lower
flux bound respectively.  Expanding this to the various cases gives
the CS relations quoted in the text.

Still following the development in \cite{lpbook}, we can read off
from \eref{eq:app1} that the dual objective function is to
minimise $w=\sum''_\alpha \ya\vimax$.  But $\ya=-\Ba=\mi$ for exchange
reactions, hence $w = \sum''_i\mi\vimax$, as quoted in the main text.

This whole approach can readily be extended to other classes of
reactions.  For example a reaction with a prescribed flux leads to the
corresponding $\Ba$ being unrestricted and a contribution $\Ba\vafix$
being added to $w$ where $\vafix$ is the fixed flux value.  As a
special case of this, a reaction which is disabled ($\vafix = 0$) has
the corresponding $\Ba$ made unrestricted in the dual problem.  As
another special case, an exchange reaction with a specified flux has a
contribution $-\mi\vafix$ added to $w$.  It follows that $\mi = -
\partial\vbio/\partial\vafix$.  This proves the shadow prices are
yield coefficients since adding an exchange reaction $\Mi\exchange$
with a fixed flux $\va = \vafix < 0$ corresponds to adding the
metabolite at a rate $|\vafix|$.

Let us finally consider the general case $\vamin\le\va\le\vamax$.
This actually specifies two constraints on the flux, and thus gives
rise to two dual variables.  One possible interpretation of the
resulting dual problem is as follows.  In addition to $\Ba\le0$ one
has $\Ba\le\sum_i\mi\Sia$ rather than a strict equality.  A double
contribution $\Ba\vamin + (\sum_i\mi\Sia - \Ba)\vamax$ is added to
$w$.  There are also two CS relations, namely $\Ba(\va-\vamin)=0$ and
$(\sum_i\mi\Sia-\Ba)(\vamax-\va)=0$.


\end{document}